\documentclass[a4paper,11pt]{article}
\usepackage{amssymb,amsmath,amsfonts}

\newtheorem{idea memo}[theorem]{Idea Memo}

\input tcilatex.tex
\begin{document}

\title{Roles of Asymptotic Condition and S-Matrix as Micro-Macro Duality in
QFT}
\author{Izumi OJIMA \\
Research Institute for Mathematical Sciences, \\
Kyoto University, Kyoto 606-8502, Japan}
\date{}
\maketitle

\begin{abstract}
Various versions of \textquotedblleft independence\textquotedblright\ are
actively inverstigated in quantum probability. In the context of
relativistic QFT, we show here that the physical origin of \textquotedblleft
independence\textquotedblright\ can be sought in the asymptotic condition
through which asymptotic fields and states exhibiting the independence
emerge from the non-independent interacting Heisenberg fields in a kind of
\textquotedblleft central limit\textquotedblright . From the algebraic
viewpoint, this condition is equivalent to the on-shell condition to pick up
free one-particle modes, which also reduces to Einstein's famous formula $%
E=mc^{2}$. A scenario to reconstruct interacting Heisenberg fields as
Micro-objects from these \textquotedblleft independent\textquotedblright
=free Macro-objects intertwined by an S-matrix as a measurable quantity is
formulated according to the Micro-Macro Duality associated with a new notion
of a \textit{cocycle of K-T operators}.
\end{abstract}

\section{Two questions: What do independence and $E=mc^{2}$ mean?}

In the theory of quantum probability, several versions of \textit{%
independence} \cite{Indep} have been formulated with interesting results as
generalizations of the bosonic tensor type. My na\"{\i}ve questions here are 
\textit{on which physical grounds} they have appeared and \textit{what
physical meanings} they have. At least for the familiar Gaussian cases
(=quasi-free states on bosonic CCR~or fermionic CAR), my partial answers in
the context of relativistic QFT will be given as follows:

1) \textit{\textbf{Emergence of independence}}\ through the \textit{\textbf{%
asymptotic condition }} $\varphi _{H}(x)\underset{x^{0}=t\rightarrow \mp
\infty }{\rightarrow }\phi ^{in/out}(x)$:

From \textit{non-independent} interacting Heisenberg fields $\varphi _{H}$,
the \textit{asymptotic fields\textbf{\ }}$\phi ^{as}=\phi ^{in/out}$ and%
\textit{\ asymptotic states} generated by $\phi ^{as}$ from the vacuum 
\textit{satisfying the \textbf{independence}} arise through the \textit{%
asymptotic condition }\cite{LSZ}, $\varphi _{H}(x)\underset{%
x^{0}=t\rightarrow \mp \infty }{\rightarrow }\phi ^{in/out}(x)$, which can
be interpreted as a sort of \textquotedblleft central
limit\textquotedblright\ theorem. The conceptual meaning of this
\textquotedblleft central limit\textquotedblright\ theorem\ can be placed in
the \textquotedblleft \textit{\textbf{Micro-Macro Duality}}%
\textquotedblright \textit{\textbf{\ }}\cite{MicMac}\textit{\textbf{\ }}in
QFT as follows: 
\begin{equation*}
\begin{array}{ccc}
\text{Micro} &  & \text{Macro} \\ 
\varphi _{H}\text{ {\small : }generic} & 
\begin{array}{c}
\text{asymp.cond.} \\ 
\rightleftarrows \\ 
\text{{\small GLZ expansion of} }\varphi _{H}\text{ {\small in} }\phi ^{as}%
\end{array}
& \phi ^{as}\text{{\small : }universal} \\ 
\begin{array}{c}
{\small (\square +m}^{2}{\small )\varphi }_{H}{\small =J}_{H} \\ 
\Longleftrightarrow {\small \varphi }_{H}{\small =\Delta }_{ret}{\small \ast
J}_{H}{\small +\phi }^{in}%
\end{array}
&  & 
\begin{array}{c}
p^{2}=m^{2}\Longleftrightarrow \\ 
{\small (\square +m}^{2}{\small )\phi }^{as}{\small =0}%
\end{array}%
\end{array}%
\end{equation*}

(NB: There exists another local-net version of \textquotedblleft statistical
independence\textquotedblright\ based on the so-called \textit{nuclearity
condition} \cite{Nucl} in Algebraic QFT, which should not be confused with
the present version.)

2) \textquotedblleft Units\textquotedblright\ of \textit{independence}
identified with \textit{particles} specified by Einstein's famous formula $%
E=mc^{2}$ ($\Longrightarrow $ Sec.2):

It is worth noting that this famous formula $E=mc^{2}$ is meaningful only
for asymptotic fields/ states as the \textit{on-shell condition} $%
p^{2}=p_{\mu }p^{\mu }=m^{2}$ to extract\textit{\textbf{\ }}particles as 
\textit{independent = free = non-interacting\textbf{\ }}entities from the
interacting Heisenberg fields; in contrast, the latter do not satisfy this
formula because of the interactions. The former, asymptotic fields and
states, serve as the \textit{vocabulary for describing state changes\textbf{%
\ }}taking place in the scattering processes in such a form as the state
transitions described by the S-matrix $\langle \beta ,out|\alpha ,in\rangle
=\langle \beta |S|\alpha \rangle $ from an \textit{asymptotic incoming state%
\textbf{\ }}(\textit{in-state}, for short)\textit{\textbf{\ }}$|\alpha
,in\rangle $ to an \textit{outgoing state} (\textit{out-state}, for short) $%
|\beta ,out\rangle $. In the absence of interactions, however, the \textit{%
on-shell\textbf{\ }}asymptotic fields $\phi ^{as}$ can\textit{\textbf{not}}
by themselves ignite scattering processes, which necessitates the use of 
\textit{off-shell interacting Heisenberg fields\textbf{\ }}$\varphi _{H}$.

3) The logical basis of the \textit{asymptotic condition} is to be found in
the \textit{cluster property }\cite{StrWi}: 
\begin{equation*}
\langle \Omega |A\alpha _{\vec{x}}(B)\Omega \rangle \underset{\vec{x}%
\rightarrow \infty }{\rightarrow }\langle \Omega |A\Omega \rangle \langle
\Omega |B\Omega \rangle ,
\end{equation*}%
following from the ergodicity valid for a unique vacuum vector $\Omega $
invariant under spacetime translations $U(x)$ and from the assumption of the
local commutativity ($\Longrightarrow $ Sec.3). In this context, asymptotic
fields $\phi ^{as}$~can be understood as quantities to\textit{\ }materialize 
\textit{kinematically} this factorization (= independence)\ of correlations
without taking limit and they can be decomposed into creation and
annihilation operators $a(\vec{p}).a^{\ast }(\vec{q})$ which represent 
\textit{infinite number of conserved quantities}.

4)\ \textit{\textbf{Universality}} of \textquotedblleft central
limit\textquotedblright\ due to Haag-GLZ expansion ($\Longrightarrow $
Sec.5):\ 

From the asymptotic condition, the Yang-Feldman equation $\varphi
_{H}=\Delta _{ret}\ast J_{H}+\phi ^{in}$ can be derived with the Heisenberg
source current $J_{H}\ $formally defined by $J_{H}=(\square +m^{2})\varphi
_{H}$ \cite{Bog}. The analogue of the \textquotedblleft Fock
expansion\textquotedblright\ \cite{Oba} in WNA can be found in the Haag-GLZ
expansion \cite{GLZ, IO89}, 
\begin{equation*}
SA=:(\omega _{0}\otimes id)(T(A\otimes 1)\exp (iJ_{H}\otimes \phi ^{in})):,
\end{equation*}%
where $\omega _{0}=\langle \Omega |\cdots \Omega \rangle $ is the vacuum
state and $S=:(\omega _{0}\otimes id)(T(\exp (iJ_{H}\otimes \phi ^{as})):$
is the \textit{S-matrix}. By this formula, the Heisenberg observables $A$
depending on $\varphi _{H}$ can be expressed in terms of the asymptotic
fields $\phi ^{as}$.

\section{What does $E=mc^{2}$ mean?}

While \textit{Einstein's famous equality\textbf{\ }}$E=mc^{2}$ between
energy and mass has been regarded as one of the most fundamental
consequences of the relativity theory, however, its actual content is simply
the \textquotedblleft \textit{\textbf{on-shell condition}}%
\textquotedblright\ to pick up \textit{1-particle modes}, meaningful only
for the \textit{independent = free = non-interacting\textbf{\ }}asymptotic
fields/states. In fact, taking $m$ as \textquotedblleft moving
mass\textquotedblright\ $m=\dfrac{m_{0}}{\sqrt{1-v^{2}/c^{2}}}$, we have 
\begin{eqnarray*}
&&E=mc^{2}=\dfrac{m_{0}}{\sqrt{1-v^{2}/c^{2}}}c^{2}\text{ \ } \\
&\Longrightarrow &\text{ \ }(m_{0}c)^{2}=(\frac{E}{c})^{2}(1-v^{2}/c^{2})=(%
\frac{E}{c})^{2}-\left( \dfrac{m_{0}}{\sqrt{1-v^{2}/c^{2}}}\vec{v}\right)
^{2}=(\frac{E}{c})^{2}-\left( \vec{p}\right) ^{2} \\
&\Longrightarrow &p^{2}=p_{\mu }p^{\mu }=(m_{0}c)^{2}
\end{eqnarray*}%
where $\dfrac{m_{0}}{\sqrt{1-v^{2}/c^{2}}}\vec{v}=:\vec{p}$ is the
relativistic 3-momentum and $p^{\mu }=(\dfrac{E}{c},\vec{p})$ is the
4-mementum. The meaning of the above equality $p^{2}=p_{\mu }p^{\mu }=(%
\dfrac{E}{c})^{2}-\left( \vec{p}\right) ^{2}=(m_{0}c)^{2}$ can be understood
as follows:

i) It is just the \textit{mass-shell\textbf{\ }}(or, \textit{on-shell})%
\textit{\textbf{\ }condition} to characterize a mass hyperboloid in the $p$%
-space of 4-momenta $p_{\mu }\in \mathbb{\hat{R}}^{4}$ carried by the free
1-particle states with a rest mass $m_{0}$. The spacetime geometry inherent
to the special relativity is controlled by the Poincar\'{e} group $\mathcal{P%
}_{+}^{\uparrow }=\mathbb{R}^{4}\rtimes L_{+}^{\uparrow }$ (or, its
universal covering $\widetilde{\mathcal{P}_{+}^{\uparrow }}=\mathbb{R}%
^{4}\rtimes SL(2,\mathbb{C})$) defined by the semi-direct product of
spacetime translation group $\mathbb{R}^{4}$ and the orthochronous proper
Lorentz group $L_{+}^{\uparrow }:=\{\Lambda =(\Lambda _{\nu }^{\mu });$ $%
\Lambda x\cdot \Lambda y=x\cdot y,\Lambda _{0}^{0}>0,\det (\Lambda )=+1\}$
consisting of homogeneous Lorentz transformations $\Lambda =(\Lambda _{\nu
}^{\mu })\in SO(1,3)$ leaving the Minkowski metric $\eta (x,y):=x\cdot
y=x^{0}y^{0}-\vec{x}\cdot \vec{y}$ invariant, $\Lambda ^{T}\eta \Lambda
=\eta $, without changing the time direction $\Lambda _{0}^{0}>0$. In
Wigner's construction \cite{Bog} of irreducible unitary representations of $%
\mathcal{P}_{+}^{\uparrow }$ or $\widetilde{\mathcal{P}_{+}^{\uparrow }}$,
four orbit families, $p^{2}\overset{>}{\underset{<}{=}}0$ and $p_{\mu }=0$,
appear: $p^{2}=m_{0}^{2}>0$ corresponds to a massive particle with a rest
mass $m_{0}$, $p^{2}=0$ to massless particles, $p^{2}<0$ to (unphysical)
\textquotedblleft tachyons\textquotedblright\ (with an imaginary mass) and
the last one to the vacuum, each of which is induced from one of the
corresponding \textquotedblleft little groups\textquotedblright\ ($%
SU(2),E(2),SU(1,1)$ and $L_{+}^{\uparrow }$).

ii)\ Through the \textquotedblleft first quantization\textquotedblright\ $%
p_{\mu }\rightarrow i\hbar \partial _{\mu }=i\hbar (\dfrac{1}{c}\dfrac{%
\partial }{\partial t},\vec{\nabla})$, the \textit{Klein-Gordon equation} $%
[\hbar ^{2}\partial _{\mu }\partial ^{\mu }+(m_{0}c)^{2}]\phi (x)=0~$%
describes a \textit{free scalar field}~$\phi (x)$~with rest mass $m_{0}$.

iii) The existence of \textit{positive/ negative energy} solutions $E=\pm 
\sqrt{\left( \vec{p}c\right) ^{2}+(m_{0}c^{2})^{2}}$ of $(\frac{E}{c}%
)^{2}-\left( \vec{p}\right) ^{2}=m_{0}^{2}c^{2}$ is related with creation
and annihilation operators, particle-antiparticle pairs, time reversal 
\textit{T\textbf{\ }}and \textit{PCT invariance}.

Thus, the famous equivalence $E=mc^{2}$ between energy $E$ and mass $m$
gives only \textit{partial} information for dynamical descriptions of
relativistic quantum fields, with \textit{off-shell\textbf{\ }}apects being
neglected in spite of their vital importance for non-trivial scattering
processes, particle decays and productions, etc., etc.

\section{Free= \textit{\textbf{independent}} vs. interacting= \textit{%
\textbf{non-independent}}}

A free quantum field $\phi (x)$ as the quantized solution of Klein-Gordon
equation $(\square +m^{2})\phi =0$ describes \textquotedblleft particle
pictures\textquotedblright\ in terms of creation and annihilation operators $%
a(f),a^{\ast }(f)$ defined as follows: 
\begin{eqnarray*}
\phi (x) &=&\int \dfrac{d^{3}p}{\sqrt{(2\pi )^{3}2\omega _{\vec{p}}}}(a(\vec{%
p})\exp (-ip_{\mu }x^{\mu })+h.c.), \\
a(f) &:&=i\int \overline{f(x)}\overleftrightarrow{\partial _{0}}\phi
(x)d^{3}x=\int \overline{\tilde{f}(\vec{p})}a(\vec{p})d^{3}p, \\
a^{\ast }(f) &:&=i\int \phi (x)\overleftrightarrow{\partial _{0}}%
f(x)d^{3}x=\int a^{\ast }(\vec{p})\tilde{f}(\vec{p})d^{3}p=[a(f)]^{\ast } \\
\lbrack a(f),a^{\ast }(g)] &=&\int \overline{\tilde{f}(\vec{p})}\tilde{g}(%
\vec{p})d^{3}p=\langle \tilde{f},\tilde{g}\rangle , \\
\lbrack \phi (x),\phi (y)] &=&\int \dfrac{d^{4}p}{(2\pi )^{3}}\varepsilon
(p^{0})\delta (p^{2}-m^{2})\exp (-ip(x-y))=:i\Delta (x-y;m^{2}),
\end{eqnarray*}%
with $\omega _{\vec{p}}:=\sqrt{\vec{p}^{2}+m^{2}}$ in the \textquotedblleft
natural unit system\textquotedblright\ with $\hbar =c=1$ (rest mass $m_{0}$
is denoted by $m$, henceforth).

Although free quantum fields $\phi (x)$ with $a^{\ast }(\vec{p}),a(\vec{p})$
are customarily believed to be sufficient entities for describing \textit{%
wave-particle dualism\textbf{\ }}inherent in elementary particles, the
perpetual creation and annihilation processes of particles require \textit{%
interactions} among elementary particles, which is not consistent with the
linearity of free field equations. Concerning this point, the famous theorem
of Haag \cite{StrWi, Bog} has been taken as a kind of \textquotedblleft
no-go theorem\textquotedblright\ for the theoretical description of the
interactions: \textit{Haag theorem}: Poincar\'{e} (or even,
Galilei)-covariant quantum fields related to free fields by a unitary
transformation are only free fields. Owing to this theorem, it is
meaningless to formulate the interacting Heisenberg fields by means of a
unitary transformation of free fields as is commonly done in perturbative
approaches. This is in sharp contrast to the qunatum systems with \textit{%
finite} degrees of freedom.

On the other hand, to describe relativistic scattering processes of
elementary particles in a satisfactory way, we need inevitably the following
ingredients: Poincar\'{e}-covariant quantum fields/ their interactions/ free
asymptotic fields and states. Here, free fields are indispensable as the
vocabulary for the description of scattering processes, where an initial
state with incoming free particles is changed into a final one with outgoing
particles.

Giving up the idea to connect directly Heisenberg and asymptotic fields (as
is forbidden by the Haag theorem), we consider the mutual relations between%
\textit{\textbf{\ }two free asymptotic fields},\textit{\textbf{\ }}$\phi
^{in}(x)$ and $\phi ^{out}(x)$ in terms of the \textit{unitary S-matrix%
\textbf{\ }}$S$ to control the \textit{basis change\textbf{\ }}between the
in-state basis $|\alpha ,in\rangle $ and the out-state basis $|\beta
,out\rangle $: 
\begin{equation*}
|\alpha ,in\rangle =\underset{\beta }{\sum }|\beta ,out\rangle S_{\beta
,\alpha }\text{ \ \ \ with }S_{\beta ,\alpha }:=\langle \beta ,out|\alpha
,in\rangle =\langle \beta |S|\alpha \rangle ,
\end{equation*}%
The schematic picture can be summarized as follows: 
\begin{equation*}
\begin{array}{c||c|c|c}
& \text{in} &  & \text{out} \\ \hline\hline
\begin{array}{c}
\text{asymptotic~} \\ 
\text{fields}%
\end{array}
& 
\begin{array}{c}
{\small Ad\Theta }^{in}{\small \curvearrowright }\text{\ \ \ \ } \\ 
\text{ \ \ \ \ \ \ }\phi ^{in}(x)%
\end{array}
& \overset{AdS^{-1}\text{ \& }Ad\Theta }{\underset{AdS\text{ \& }Ad\Theta }{%
\rightleftarrows }} & 
\begin{array}{c}
\text{\ \ \ \ \ }{\small \curvearrowleft Ad\Theta }^{out} \\ 
\phi ^{out}(x)\text{ \ \ \ \ \ \ }%
\end{array}
\\ \hline
& \text{GLZ}\searrow \nwarrow t\rightarrow -\infty & 
\begin{array}{c}
\text{asymp.cond.}\uparrow \\ 
\downarrow \text{GLZ formula}%
\end{array}
& t\rightarrow +\infty \nearrow \swarrow \text{GLZ } \\ \hline
\begin{array}{c}
\text{Heisenberg~} \\ 
\text{fields}%
\end{array}
&  & 
\begin{array}{c}
{\small Ad\Theta \curvearrowright }\text{ \ \ } \\ 
\text{ \ \ \ \ \ }\varphi _{H}(x)%
\end{array}
& 
\end{array}%
\end{equation*}

To treat Heisenberg fields~$\varphi _{H}(x)$, we recapitulate briefly the
essence of Wightman axioms for relativistic quantum fields \cite{StrWi, Bog}
(in the vacuum representation $(\mathcal{P},\mathfrak{H},U,\Omega )$) in the
form of relativistic covariance, local commutativity, cyclicity or
ergodicity of vacuum state and spectral condition:

a) [Heisenberg fields] as operator-valued distributions $\mathcal{D}(\mathbb{%
R}^{4})\ni f\longmapsto \varphi _{H}^{i}(f)$ with values being (unbounded)
closable operators acting on a Hilbert space $\mathfrak{H}$ are defined on
the 4-dimensional Minkowski spacetime $(\mathbb{R}^{4},\eta )$, where $\eta $%
~is the Minkowski metric: $\eta (x,y):=x\cdot y=x^{0}y^{0}-\vec{x}\cdot \vec{%
y}$.

b) [Relativistic covariance]: a local net $\mathcal{P}:\mathcal{K}\ni 
\mathcal{O}\longmapsto \mathcal{P}(\mathcal{O})$ of *-algebras $\mathcal{P}(%
\mathcal{O})$ generated by local fields $\varphi _{H}^{i}(f)=\int \varphi
_{H}^{i}(x)f(x)d^{4}x$ with $f\in \mathcal{D}(\mathcal{O})$ and their
polynomials defined on the net $\mathcal{K}$ of double cones $\mathcal{O}=%
\mathcal{O}_{a,b}=(a+V_{+})\cap (b-V_{+})$ (with the forward lightcone $%
V_{+} $) constitute a non-commutative covariant dynamical system, 
\begin{eqnarray*}
\alpha _{a,\Lambda }(\varphi _{H}^{i}(x)) &=&U(a,\Lambda )\varphi
_{H}^{i}(x)U(a,\Lambda )^{-1} \\
&=&s(\Lambda )_{j}^{i}\varphi _{H}^{i}(\Lambda ^{-1}(x-a)), \\
\alpha _{a,\Lambda }(\mathcal{P}(\mathcal{O})) &=&\mathcal{P}(\Lambda 
\mathcal{O}+a),
\end{eqnarray*}%
under the action $\alpha $, $\mathcal{P}_{+}^{\uparrow }\ni (a,\Lambda
)\longmapsto \alpha _{a,\Lambda }\in Aut(\mathcal{P}(\mathbb{R}^{4}))$, of
Poincar\'{e} group $\mathcal{P}_{+}^{\uparrow }$ (or its covering $%
\widetilde{\mathcal{P}_{+}^{\uparrow }}$ ) and $\mathcal{P}_{+}^{\uparrow
}\ni (a,\Lambda )\longmapsto U(a,\Lambda )\in \mathcal{U}(\mathfrak{H})$ is
its unitary representation on $\mathfrak{H}$, and $s(\Lambda )_{j}^{i}$~is a
finite-dimensional representation of Lorentz group $L_{+}^{\uparrow }$
associated with each field multiplet $(\varphi _{H}^{i}(x))_{i}$.

c) [Local commutaitivity]: the absence of propagation of physical effects
exceeding the light velocity due to Einstein causality, implies the local
commutativity of Heisenberg fields $\varphi _{H}^{i}(f)$: 
\begin{equation*}
\lbrack \varphi _{H}^{i}(f_{1}),\varphi _{H}^{j}(f_{2})]=0\text{ \ \ \ if }%
(suppf_{1}){\huge \times }(suppf_{2})
\end{equation*}%
where$~\mathcal{O}_{1}{\huge \times }\mathcal{O}_{2}$ means that any pair of
points $x\in \mathcal{O}_{1}{\LARGE ,}y\in \mathcal{O}_{2}$ are spacelike
separated: $(x-y)^{2}<0$.

Remark: By this condition, the Fourier transform of Wightman functions $%
\omega _{0}(\varphi _{H}^{i_{1}}(x_{1})\cdots \varphi _{H}^{i_{r}}(x_{r}))$
as correlation functions of $\varphi _{H}^{i}$ in the vacuum state $\omega
_{0}(\cdot )=\langle \Omega |(\cdot )\Omega \rangle $ defined in the next d)
admits an analytic continuation into a holomorphic function in the complex
energy-momentum space. According to it, dispersion relations are valid.

d) [Vacuum state and spectrum condition]:

d-i) Energy-momentum spectrum $Sp(U(\mathbb{R}^{4}))$ of spacetime
translations $\mathbb{R}^{4}$ realized on $\mathfrak{H}$ is within the
forward light cone, $Sp(U(\mathbb{R}^{4}))\subset \overline{V_{+}}$ in $p$%
-space $\widehat{\mathbb{R}^{4}}$, and the lowest energy is realized by
eigenvalue $0$ of the vacuum vector $\Omega $: $U(x):=U(x,1)=\int_{p\in 
\overline{V_{+}}}\exp (ipx)dE(p)$; \ $U(x)\Omega =\Omega $

Remark: Similarly to $p$-analyticity due to local commutativity, $x$-space
analyticity of Wightman functions $\omega _{0}(\varphi
_{H}^{i_{1}}(x_{1})\cdots \varphi _{H}^{i_{r}}(x_{r}))$ follows from
spectrum condition, which provides powerful tools for structural analysis.

d-ii)\ Cyclicity $\overline{\mathcal{P}(\mathbb{R}^{4})\Omega }=\mathfrak{H}$
of $\Omega $ $\Longleftrightarrow $ irreducibility of $\mathcal{P}(\mathbb{R}%
^{4})$ $\Longleftrightarrow $ uniqueness of vacuum (: $U(x)\Psi =\Psi
\Longrightarrow \Psi \propto \Omega $) $\Longleftrightarrow $ cluster
property:%
\begin{equation*}
\left\vert \omega _{0}(A(x)B(y))-\omega _{0}(A)\omega _{0}(B)\right\vert
\rightarrow 0\text{\ as }(\vec{x}-\vec{y})^{2}\rightarrow \infty ,
\end{equation*}%
where $A(x):=\alpha _{x}(A)=U(x)AU(x)^{\ast },$ $B(y):=\alpha _{y}(B)$ are
the spacetime translates of local observables $A,B\in \mathcal{P}(\mathcal{O}%
)$ by $x,y\in \mathbb{R}^{4}$, respectively. This follows from partition of
unity due to spectral resolution of spacetime translations $U(x)$: 
\begin{eqnarray*}
1 &=&|\Omega \rangle \langle \Omega |\text{ }+\text{ }\sum_{i}(\text{\textit{%
1-particle singularities on mass-shell}}\mathit{\mathbf{\ }}p^{2}=m_{i}^{2})
\\
&&+(\text{absolutely continuous }p\text{-spectra})
\end{eqnarray*}

\section{Independence of asymptotic fields due to on-shell asymptotic
condition as \textquotedblleft central limit\textquotedblright\ theorem}

From the \textit{cluster property} and the local commutativity, follows the 
\textit{asymptotic condition} $\varphi _{H}(x)\underset{x^{0}=t\rightarrow
\mp \infty }{\rightarrow }\phi ^{in/out}(x)$ (as weak convergence),
according to which asymptotic fields $\phi ^{as}$~materialize \textit{%
kinematically} the factorization (= independence)\ of correlations \textit{%
without taking limits}: any $n$-point functions of $\phi ^{as}$ are
factorized into the products of two-point functions, $\omega _{0}(\phi
^{as}\phi ^{as}\cdots \phi ^{as})=\sum \omega _{0}(\phi ^{as}\phi
^{as})\cdots \omega _{0}(\phi ^{as}\phi ^{as})$, which is known as the
\textquotedblleft quasi-freeness\textquotedblright\ of the vacuum state $%
\omega _{0}$ with respect to $\phi ^{as}$, familiar in the form of
\textquotedblleft Wick theorem\textquotedblright\ or the \textit{independence%
} of Gaussian type. The creation and annihilation operators $%
a_{k},a_{k}^{\ast }$ contained in $\phi ^{as}$ constitute an \textit{%
infinite number of conserved quantities}, as they are given by the spatial
integral of current densities $i\phi ^{as}(x)\overleftrightarrow{\partial
_{\mu }}f(x)$ which are conserved, $\partial ^{\mu }[\phi ^{as}(x)%
\overleftrightarrow{\partial _{\mu }}f(x)]=\phi ^{as}(x)\overleftrightarrow{%
\square }f(x)=\phi ^{as}(x)\overleftrightarrow{(\square +m^{2})}f(x)=0$ by
the on-shell conditions: $(\square +m^{2})\phi ^{as}(x)=0=(\square
+m^{2})f(x)$.

Thus, the \textit{independence} embodied in the asymptotic fields $\phi
^{as} $ is seen to emerge from interacting Heisenberg fields $\varphi _{H}$
via the asymptotic condition as a kind of \textquotedblleft central
limit\textquotedblright\ theorem. In this context, what corresponds to
\textquotedblleft Langevin equation\textquotedblright\ can be found in the 
\textit{Yang-Feldman equation} \cite{Bog} to connect the Heisenberg field~$%
\varphi _{H}(x)$~and the asymptotic field $\phi ^{as}(x)$: 
\begin{eqnarray*}
\varphi _{H}(x) &=&\int \Delta _{ret}(x-y;m^{2})J_{H}(y)d^{4}y+\phi
^{in}(x)=[\Delta _{ret}\ast J_{H}+\phi ^{in}](x) \\
&=&\int \Delta _{adv}(x-y;m^{2})J_{H}(y)d^{4}y+\phi ^{out}(x)=[\Delta
_{adv}\ast J_{H}+\phi ^{out}](x).
\end{eqnarray*}%
where $J_{H}=(\square +m^{2})\varphi _{H}$ is the Heisenberg source current
and $\Delta _{ret/adv}(x-y;m^{2})$ : retarded/ advanced Green's functions
(i.e., principal solutions) of the Klein-Gordon equation defined by 
\begin{eqnarray*}
(\square _{x}+m^{2})\Delta _{ret/adv}(x-y;m^{2}) &=&\delta (x-y), \\
\Delta _{ret/adv}(x-y;m^{2}) &=&0\text{ \ \ \ for }x_{0}\lessgtr y_{0}.
\end{eqnarray*}%
In the Yang-Feldman equation, the asymptotic fields $\phi ^{in/out}$ and
Heisenberg source current $J_{H}$ appear, respectively, as the \textit{%
residue} and the \textit{quotient} in the division of $\varphi _{H}$ by $%
\Delta _{ret/adv}$. What is more important is that $J_{H}$ gives the \textit{%
residues at the on-shell pole} $\dfrac{1}{p^{2}-m^{2}}$ to determine matrix
elements of scattering amplitudes, as will be seen in the next formula.

\section{Mutual relations between Heisenberg fields and asymptotic fields
controlled by Micro-Macro Duality}

From the asymptotic condition and the LSZ reduction formulae \cite{LSZ}, one
can derive the Haag-GLZ formulae \cite{GLZ}~to express Heisenberg operators $%
A$ in terms of the Wick products $:\phi ^{as}\cdots \phi ^{as}:$ of
asymptotic fields:%
\begin{eqnarray*}
SA &=&:\exp (\phi ^{in}(\square +m^{2})\dfrac{\delta }{\delta J}):\omega
_{0}(T(A\exp (i\varphi _{H}J))\upharpoonright _{J=0} \\
&=&\sum_{k=0}^{\infty }\dfrac{i^{k}}{k!}\int dx_{1}\cdots \int
dx_{k}(\square _{x_{1}}+m^{2})\cdots (\square _{x_{1}}+m^{2})\omega
_{0}(T(A\varphi _{H}(x_{1})\cdots \varphi _{H}(x_{k})) \\
&&\times :\phi ^{in}(x_{1})\cdots \phi ^{in}(x_{k}):, \\
S &=&:\exp (\phi ^{as}(\square +m^{2})\dfrac{\delta }{\delta J}):\omega
_{0}(T(\exp (i\varphi _{H}J))\upharpoonright _{J=0},
\end{eqnarray*}%
which is similar to the Fock expansion formula \cite{Oba} known in
WNA.~While the formulae of this type have long been known simply as those to
expand the Heisenberg operators in the Wick products of $\phi ^{as}$, what
should be emphasized here are the following novel points:

1)\ We can reformulate these equalities into the following form \cite{IO89}: 
\begin{eqnarray*}
SA &=&:(\omega _{0}\otimes id)(T(A\otimes 1)\exp (iJ_{H}\otimes \phi
^{in})):, \\
A &=&S^{-1}:(\omega _{0}\otimes id)(T[A\otimes 1]\exp (iJ_{H}\otimes \phi
^{in})): \\
&=&:(\omega _{0}\otimes id)(T[A\otimes 1]\exp (iJ_{H}\otimes \phi
^{out}):S^{-1}, \\
S &=&:(\omega _{0}\otimes id)(T\exp (iJ_{H}\otimes \phi ^{in})):=:(\omega
_{0}\otimes id)(T\exp (iJ_{H}\otimes \phi ^{out}):.
\end{eqnarray*}%
The roles played by the system consisting of Heisenberg operators $\varphi
_{H}$ and those of asymptotic fields $\phi ^{in}$ and $\phi ^{out}$ are
clearly separated here in the \textit{\textbf{operational context}} in such
a way that the former is an \textit{unknown} target system to be detected
and analyzed by means of the latter ones functioning as \textit{\textbf{%
probe systems}}. The relevant \textit{\textbf{coupling terms }}among them
are specified by $\exp (iJ_{H}\otimes \phi ^{as})$, with the first tensor
factor to be time-ordered and the second one Wick ordered, which are
mutually in duality, as will be seen by the relation between $\dfrac{\delta 
}{i\delta \phi ^{in}(x)}$ and $\phi ^{in}(x)$ in 3).~In this connection, the
comparison to the notion of instruments in quantum measurements will be
instructive: while a basic scheme for instruments can be seen in 
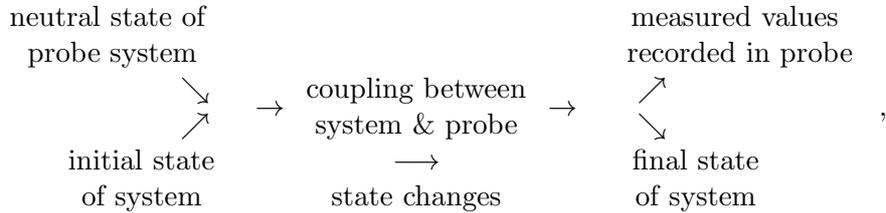
\begin{equation*}
\begin{array}{ccc}
\begin{array}{c}
\text{neutral state of } \\ 
\text{probe system}%
\end{array}
&  & 
\begin{array}{c}
\text{measured values} \\ 
\text{ recorded in probe}%
\end{array}
\\ 
\text{ \ \ \ \ \ \ \ \ \ \ \ \ \ \ \ \ }%
\begin{array}{c}
\searrow \\ 
\nearrow%
\end{array}
& \rightarrow 
\begin{array}{c}
\text{coupling between} \\ 
\text{system \& probe}%
\end{array}%
\rightarrow & 
\begin{array}{c}
\nearrow \\ 
\searrow%
\end{array}%
\text{ \ \ \ \ \ \ \ \ \ \ \ \ \ \ \ \ } \\ 
\ \ \ \ \ \ 
\begin{array}{c}
\text{initial state} \\ 
\text{of system}%
\end{array}
& 
\begin{array}{c}
\longrightarrow \\ 
\text{state changes}%
\end{array}
& 
\begin{array}{c}
\text{final state} \\ 
\text{of system}%
\end{array}%
\text{ \ \ \ \ \ \ \ }%
\end{array}%
,
\end{equation*}%
the corresponding one for scattering processes in QFT is given in sharp
contrast by 
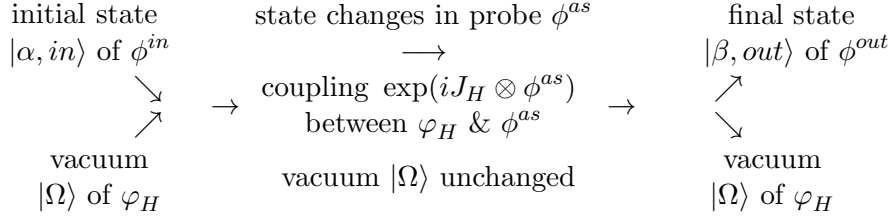
\begin{equation*}
\begin{array}{cccc}
\begin{array}{c}
\text{initial state } \\ 
|\alpha ,in\rangle \text{ of }\phi ^{in}%
\end{array}
& 
\begin{array}{c}
\text{state changes in probe }\phi ^{as} \\ 
\longrightarrow%
\end{array}
&  & 
\begin{array}{c}
\text{final state } \\ 
|\beta ,out\rangle \text{ of }\phi ^{out}%
\end{array}
\\ 
\text{ \ \ \ \ \ \ \ \ \ \ \ }%
\begin{array}{c}
\searrow \\ 
\nearrow%
\end{array}
& \rightarrow 
\begin{array}{c}
\text{coupling }\exp (iJ_{H}\otimes \phi ^{as})\text{ } \\ 
\text{between }\varphi _{H}\text{ \& }\phi ^{as}%
\end{array}%
\rightarrow &  & 
\begin{array}{c}
\nearrow \\ 
\searrow%
\end{array}%
\text{ \ \ \ \ \ \ \ \ \ \ \ \ \ } \\ 
\ \ 
\begin{array}{c}
\text{vacuum} \\ 
|\Omega \rangle \text{ of }\varphi _{H}%
\end{array}
& \text{ vacuum }|\Omega \rangle \text{ unchanged} &  & 
\begin{array}{c}
\text{vacuum} \\ 
|\Omega \rangle \text{ of }\varphi _{H}%
\end{array}%
\text{ \ \ \ \ }%
\end{array}%
.
\end{equation*}%
Perhaps, they can be unified by combining the level of the object system in
the instrument with that of the asymptotic fields $\phi ^{as}$, which
results in the successive measurement processes where the state changes
taking place at the level of asymptotic fields in scattering processes of
quantum fields are monitored by measuring such observables as the particle
momenta or spins through the instruments.

2) The mathematical and conceptual meanings of the coupling term $\exp
(iJ_{H}\otimes \phi ^{in})$: We first note by simple computation that the
on-shell condition $(\square +m^{2})\phi ^{as}=0$ for the asymptotic fields
implies the following equalities: 
\begin{eqnarray*}
J_{H}\otimes \phi ^{as} &=&(\square +m^{2})\varphi _{H}\otimes \phi
^{as}=(\square +m^{2})\varphi _{H}\otimes \phi ^{as}-\varphi _{H}\otimes
(\square +m^{2})\phi ^{as} \\
&=&\square \varphi _{H}\otimes \phi ^{as}-\varphi _{H}\otimes \square \phi
^{as}=\partial ^{\mu }\partial _{\mu }\varphi _{H}\otimes \phi ^{as}-\varphi
_{H}\otimes \partial ^{\mu }\partial _{\mu }\phi ^{as} \\
&=&\partial ^{\mu }[\partial _{\mu }\varphi _{H}\otimes \phi ^{as}-\varphi
_{H}\otimes \partial _{\mu }\phi ^{as}]=-\partial ^{\mu }[\varphi
_{H}\otimes \overset{\longleftrightarrow }{\partial _{\mu }}\phi ^{as}].
\end{eqnarray*}%
Combing this with the asymptotic condition we can further rewrite this
quantity $iJ_{H}\otimes \phi ^{in}$: 
\begin{eqnarray*}
&&i\int_{\mathbb{R}^{4}}d^{4}xJ_{H}(x)\otimes \phi ^{in}(x)=-i\int_{\mathbb{R%
}^{4}}d^{4}x\partial ^{\mu }[\varphi _{H}\otimes \overset{%
\longleftrightarrow }{\partial _{\mu }}\phi ^{in}]= -i\int_{\partial \mathbb{%
R}^{4}}dS^{\mu }[\varphi _{H}\otimes \overset{\longleftrightarrow }{\partial
_{\mu }}\phi ^{in}] \\
&&=-i\int_{x^{0}=+\infty }d^{3}x[\varphi _{H}\otimes \overset{%
\longleftrightarrow }{\partial _{0}}\phi ^{in}]+i\int_{x^{0}=-\infty
}d^{3}x[\varphi _{H}\otimes \overset{\longleftrightarrow }{\partial _{0}}%
\phi ^{in}] \\
&&=-i\int_{x^{0}=+\infty }d^{3}x[\phi ^{out}\otimes \overset{%
\longleftrightarrow }{\partial _{\mu }}\phi ^{in}]+i\int_{x^{0}=-\infty
}d^{3}x[\phi ^{in}\otimes \overset{\longleftrightarrow }{\partial _{\mu }}%
\phi ^{in}] \\
&&=-(S^{-1}\otimes 1)iQ(S\otimes 1)+iQ=-(S^{-1}\otimes 1)[iQ,S\otimes 1] \\
&&=-(S^{-1}\otimes 1)ad(iQ)(S\otimes 1),
\end{eqnarray*}%
in terms of a \textit{conserved charge} $Q$ defined by 
\begin{equation*}
iQ:=i\int d^{3}x[\phi ^{in}\otimes \overset{\longleftrightarrow }{\partial
_{\mu }}\phi ^{in}]=\sum_{k}[(a_{k}^{in})^{\ast }\otimes
a_{k}^{in}-a_{k}^{in}\otimes (a_{k}^{in})^{\ast }].
\end{equation*}%
It is remarkable that $:\exp (iQ):$ (with the Wick product $:\cdots :$ \ to
be applied to the second tensor factor) is the operator to create a \textit{%
coherent state }(or,\textit{\ exponential vector})\textit{\ }$:\exp
(iQ)|\Omega \rangle :$ from the vacuum $|\Omega \rangle $ with the
Wick-ordered \textit{commutative} parameters $a_{k}^{in}$. In the context of
WNA, these quantities will correspond to the \textit{U-functionals} \cite%
{Hida} constituting a commutative algebra with respect to the Wick product.

3) Applying the above formula $A=S^{-1}:(\omega _{0}\otimes id)(T[A\otimes
1]\exp (iJ_{H}\otimes \phi ^{in})):$ to the Heisenberg source current $%
J_{H}=A$, we reproduce the expression 
\begin{equation*}
(\square +m^{2})\varphi _{H}=J_{H}=S^{-1}:(\omega _{0}\otimes
id)(T[J_{H}\otimes 1]\exp (iJ_{H}\otimes \phi ^{in})):=S^{-1}\dfrac{\delta }{%
i\delta \phi ^{in}(x)}S,
\end{equation*}%
once obtained by Bogoliubov, Medvedev and Polivanov \cite{Bog}. Through the
above relations, the asymptotic condition $\varphi _{H}\underset{%
t\rightarrow \mp \infty }{\rightarrow }\phi ^{in/out}$ to be identified with
the \textit{on-shell condtion }$\square +m^{2}=0=m^{2}-p^{2}$ is seen to
play the essential role from the algebraic viewpoint: it extracts the
asymptotic fields $\phi ^{in/out}$ from the algebra of interacting
Heisenberg fields $\varphi _{H}$ as the \textit{\textbf{fixed points }under
the Lie subgroup} $\Gamma $ generated by $\phi ^{in}$ or $Q$ in the
infinite-dimensional Heisenberg Lie group generated by $\phi ^{in}$ and $%
\dfrac{\delta }{i\delta \phi ^{in}(x)}$ with $\phi ^{in}$ taken as an
element belonging to the above commutative algebra with the Wick product.
Under the action of $\Gamma $, $a_{k}^{in}$ and $(a_{k}^{in})^{\ast }$ are
infinitely many\ conserved charges (to characterize the \textquotedblleft
integrability\textquotedblright\ of the probe system consisting of $\phi
^{as}$). Thus the group $\Gamma $ generated by (the Wick-ordered) $%
a_{k}^{as} $ and $(a_{k}^{as})^{\ast }$ characterizes the aspects of the
symmetry associated with the macroscopic on-shell situation described by $%
\phi ^{as}$. The essential features here show the strong similarity to that
of WNA if the functional derivatives $\dfrac{\delta }{i\delta \phi ^{in}(x)}$
are put in parallel with the Hida derivatives \cite{Hida}.

4) \textit{\textbf{Breakdown of invariance}} under $\Gamma $ due to the
interactions: The symmetry $\Gamma $ preserved in the probe systems
consisting of $\phi ^{as}$ is broken in the total system containing the
interacting Heisenberg fields $\varphi _{H}$, essentially due to the
presence of $\dfrac{\delta }{i\delta \phi ^{in}(x)}$ or the Heisenberg
source current $J_{H}(x)=S^{-1}\dfrac{\delta }{i\delta \phi ^{in}(x)}%
S=(\square +m^{2})\varphi _{H}(x)\neq 0$, which introduces the effects
coming from the \textit{\textbf{off-shell}} aspects of the theory. The
non-trivial existence of the coupling term $iJ_{H}\otimes \phi
^{in}=-(S^{-1}\otimes 1)[iQ,S\otimes 1]$ is equivalent to the non-triviality 
$S\neq 1$ of the S-matrix which implies the difference between $\phi ^{in}$
and $\phi ^{out}$ due to $[iQ,S\otimes 1]\neq 0$. It is interesting to note
the parallelism of this situation with the mutual relation between the first
and second laws in Newtonian mechanics: since the interaction term due to
a(n external) force is switched off in the stage of the first law, the
system enjoys a symmetry to conserve the momentum (or velocity), similarly
to the above constancy of $\phi ^{in}$ or $Q$. At the stage of the second
law, the symmetry inherent to the first law is \textit{\textbf{broken}} by
introducing the (external) force $F$, which causes the \textit{\textbf{state
changes}} described by the \textit{changes} of the momentum $p$ according to
the second law, $dp/dt=F$, in the essential use of the vocabulary provided
by the first law. The state changes in QFT due to $J_{H}=(\square
+m^{2})\varphi _{H}\neq 0$ is similarly described by the very breaking term $%
iJ_{H}\otimes \phi ^{in}=-(S^{-1}\otimes 1)[iQ,S\otimes 1]$ in terms of the
non-trivial S-matrix $S\neq 1$. At first sight, what is conserved, momentum $%
p$ and $\phi ^{as}$ (contrasted with the \textquotedblleft momentum
variables\textquotedblright\ $\dfrac{\delta }{i\delta \phi ^{as}}$), seems
to be opposite between the Newtonian and the QFT cases, but actually it is
not the case, because a particle picture with conserved $\vec{p}$ in the
Newtonian first law is embedded also in $\phi ^{as}$ through the creation
and annihilation operators, $a(\vec{p})^{\ast }$ and $a(\vec{p})$.

5) \textit{\textbf{Reconstruction}} of $\varphi _{H}$ from $\phi ^{in/out}$
intertwined by $S$: The essence of the GLZ-Fock expansion of the Heisenberg
fields $\varphi _{H}$ or $A$ in terms of the asymptotic fields $\phi ^{as}$
can be seen in the \textquotedblleft inverse problem\textquotedblright\ to 
\textit{reconstruct }the former from its fixed-point subalgebra(s) $%
\{\varphi _{H}\}^{\Gamma }=\{\phi ^{in}\}$ or $\{\phi ^{out}\}$ through the 
\textit{\textbf{co-action}} of $\Gamma $ (which is in parallel to the
so-called \textquotedblleft inverse scattering method\textquotedblright\ in
quantum mechanics to determine a potential term responsible for scattering
processes from the scattering data). This kind of \textit{duality }relation
between $\varphi _{H}$ and $\phi ^{as}$ ensures the \textit{universality} of
the asymptotic fields $\phi ^{as}$ in spite of its speciality as
statistically independent free objects. What is conceptually more important
is such a possibility to re-construct an interacting theory of relativistic
quantum fields $\varphi _{H}$ from the knowledge of an S-matrix $S$
intertwining the asymptotic in- and out-fields $\phi ^{in/out}$: $\phi ^{in}%
\underset{AdS}{\overset{AdS^{-1}}{\rightleftarrows }}\phi ^{out}$. The
crucial ingredient, $iJ_{H}\otimes \phi ^{in}=-(S^{-1}\otimes 1)[iQ,S\otimes
1]$, as a coupling term in $:(\omega _{0}\otimes id)(T((A\otimes 1)\exp
(iJ_{H}\otimes \phi ^{in})):$ $=SA$ can be determined, at least in its
integrated form, by the knowledge of $\phi ^{as}$ and $S$, the former of
which is easily constructible as free objects. The latter one is the highly
non-trivial object to be determined phenomenologically from the experimental
data of particle scattering processes to within certain limits of
exactitude. Once the functional dependence of $S$ on $\phi ^{in}$ is
specified, the formula $J_{H}(x)=S^{-1}\dfrac{\delta }{i\delta \phi ^{in}(x)}%
S$ allows a local quantity$J_{H}(x)$ to be determined. Therefore, the whole
scheme to control the mutual relations between $\varphi _{H}$ (: Micro) and $%
\phi ^{as}$ (: Macro) can be understood in the context of \textquotedblleft 
\textit{\textbf{Micro-Macro duality}}\textquotedblright\ \cite{MicMac} via
the Fourier-Galois duality between the fixed-point subalgebra and the
recovery of the total algebra as the Galois extension. A \textit{new feature}
found here is the problem related with the appearance of \textit{two} fixed
point subalgebras, $\{\phi ^{in}\}$ and $\{\phi ^{out}\}$, which are
mutually equivalent by the intertwining actions of the S-matrix, $Ad(S)$ and 
$Ad(S^{-1})$:%
\begin{equation*}
S\phi ^{out}S^{-1}=\phi ^{in},\text{ \ \ \ }S^{-1}\phi ^{in}S=\phi ^{out}.
\end{equation*}%
In this connection, the coupling terms $:T\exp (iJ_{H}\otimes \phi ^{as}):$
\ cannot be directly regarded as a Kac-Takesaki operator but should be
interpreted as a kind of a \textit{\textbf{cocycle}} between two such. This
aspects will further be elaborated.

6) PCT invariance \& Borchers classes: While the (weak) local commutativity
is inevitably violated between $\varphi _{H}$ and $\phi ^{in/out}$ and
between $\phi ^{in}$ and $\phi ^{out}$, the fields $\varphi _{H}$, $\phi
^{in}$ and $\phi ^{out}$ enjoy the local commutativity within each system.
Then the vacuum $\omega _{0}$ is invariant, $\omega _{0}\circ \theta =\omega
_{0}=\omega _{0}\circ \theta ^{as}$, under the PCT\ transformations $\theta
,\theta ^{as}$ given by $\theta (\varphi _{H}(x))=\gamma \varphi
_{H}(-x)^{\ast }$, $\theta ^{as}(\phi ^{as}(x))=\gamma \phi ^{as}(-x)^{\ast
} $ (with $\gamma \in \mathbb{T}$), and hence, $\theta ,\theta ^{as}$ are
implemented, respectively, by anti-unitary PCT$\ $operators $\Theta $ and $%
\Theta ^{as}$ s.t. $\theta (\varphi _{H}(x))=\Theta \varphi _{H}(x)\Theta $, 
$\theta ^{as}(\phi ^{as}(x))=\Theta ^{as}\phi ^{as}(x)\Theta ^{as}$ and $%
\Theta \Omega =\Theta ^{as}\Omega =\Omega $ \cite{StrWi, Bog}. Then $S\phi
^{out}(x)S^{-1}=\phi ^{in}(x)=\Theta \gamma ^{-1}\phi ^{out}(-x)^{\ast
}\Theta =\Theta \Theta ^{out}\phi ^{out}(x)\Theta ^{out}\Theta $ implies 
\begin{eqnarray*}
S &=&\Theta ^{in}\Theta =\Theta \Theta ^{out},\text{ \ \ \ }S\Theta
^{out}=\Theta =\Theta ^{in}S, \\
S\Theta &=&\Theta ^{in}=\Theta \Theta ^{out}\Theta =\Theta S,
\end{eqnarray*}%
under the assumption of asymptotic completeness. These relations exhibit
more detailed structures of $S$ in terms of PCT operators. Thus, quantum
fields with\textit{\ the same PCT operator\textbf{\ }}$\Theta $\ have the
same S-matrix $S=\Theta ^{in}\Theta =\Theta \Theta ^{out}$: this explains
the \textquotedblleft ambiguities\textquotedblright\ in the choice of
Heisenberg fields interpolating asymptotic fields $\phi ^{in/out}$ connected
by a given S-matrix $S$ in such a form as the Borchers classes \cite{Bog}\
characterized by the relative local commutativity to share the same PCT
operator $\Theta $. The consideratoins on this aspect will be crucial for
discussing the above points of 2) -- 5).

7) For physical applications in the scheme of the above 1), it would be
interesting to utilze the above formula $SA=:(\omega _{0}\otimes
id)(T(A\otimes 1)\exp (iJ_{H}\otimes \phi ^{in})):$ \ with a specific choice
of $A$ such as the electromagnetic current $A=J_{\mu }(x)$\ to analyze the
processes to measure the form factor $\langle \beta ,out|J_{\mu }(x)|\alpha
,in\rangle =\langle \beta |SJ_{\mu }(x)|\alpha \rangle $, or, in more
general contexts of \textquotedblleft weak values\textquotedblright\ \cite%
{weak} $\langle \beta ,out|A|\alpha ,in\rangle =\langle \beta |SA|\alpha
\rangle $, as suggested by Hosoya and Shikano.

The author would like to thank Profs. M. Ohya and N. Watanabe for their
invitation to QBIC2009. He is very grateful to Profs. T. Hida and Si Si and
to Prof. M. Ozawa for navigating him to the important apsects, respectively,
of WNA and of the instrument. In this connection, he has been so much
encouraged by Mr. T. Shimizu, without whose deep insights the bridge between
WNA and QFT could not have been properly appreciated. He thanks also Prof.
A. Hosoya and Mr. Y. Shikano very much for their encourgements, discussions
on his attempts to reformulate the traditional QFT along the above line and
their suggestion to examine \textquotedblleft weak values\textquotedblright
. Last but not least, he is very grateful to all the members of the regular
seminar on Math. Phys. at RIMS, Prof. S. Tanimura, Messrs. H. Andou, R.
Harada, T. Hasebe, K. Okamura and H. Saigo, for their valuable comments and
questions.

\end{document}